\documentstyle[12pt,epsfig,here]{article}
\newcommand{\bc}{\begin{center}}
\newcommand{\ec}{\end{center}}
\newcommand{\br}{\begin{flushright}}
\newcommand{\er}{\end{flushright}}
\newcommand{\bl}{\begin{flushleft}}
\newcommand{\el}{\end{flushleft}}
\newcommand{\be}{\begin{eqnarray}}
\newcommand{\ee}{\end{eqnarray}}

\begin{document}
\thispagestyle{empty}

\vbox{\vglue 0.5cm}

\bc
{\Large \bf An Insight into the Anatomy of Electro-Gravitational 
Interactions }\footnote{Talk  
given at the Seventh International Wigner Symposium,
24 - 29 August 2001.  University of Maryland, College Park, Maryland,
 U.S.A.}
\ec
\vspace{0.1cm}
\bc
{\bf Feodor F. Tikhonin} \\
\vspace{0.1cm}
{\it Institute for High Energy Physics  \\
 Protvino, Moscow Region 142284, RUSSIA \\
 e-mail tikhonin@mx.ihep.su}
\ec
\vspace{0.1cm}



\bc
{\bf Abstract}
\ec

  An interplay between electromagnetic and gravitational interactions 
is studied with particular emphasis on the particles mass dependence 
of amplitudes.  The cancellations between diagrams 
due to the gauge invarinace are explicitly demonstrated.

\sloppy
\begin{flushright}
\parbox{241pt}
                  { {\it    "No," replied Margarita, "what really puzzles 
                           me is where you have found the space for all 
                           this". With a wave of her hand Margarita 
                           emphasised the wastness of the hall they were in.
                           Koroviev smiled sweetly, wrinkling his nose. 
                           "Easy!" he replied. "For anyone who knows how to 
                           handle the {\bf fifth dimension} it's no problem to 
                           expand any place to  whatever size you please. 
                           No, dear lady, I will say more - to the devil 
                           knows what size.} \\
                           M.A.Bulgakov, "The Master and Margarita" \cite{1}}

\end{flushright}

\section{Introduction}
   The objective of this simple consideration is to attract slightly
more careful attention to the interplay between electromagnetic 
and gravitational interactions, than usually is required for cross
sections calculation.  Total cross section for the typical elementary
electrogravitational process like that considered in this note,
$ \ell^+ + \ell^-\rightarrow \gamma + $ graviton $ \equiv \cal G ,$ 
is constant proportional to $\alpha \cdot G_N$ with $\alpha \simeq
1/128$ and $G_N \simeq 2.6 \cdot 10^{-20} yb,$ while typical elementary
cross section for electroweak process (with gauge boson instead of $\cal G$)
is of the order $\alpha^2/s,$ where $s$ is the energy squared of collision.
It follows, that both cross sections should equate at energies of the order
$\sqrt{s} \simeq 10^{18}$ GeV. This huge gap is the reflection of
the hierarchy problem, one of the ways how to circumvent this problem was
proposed recently \cite{2} and outlined in a cursory manner below.

   First attempt of unified description of Electromagnetism and 
Gravity was undertaken by Theodor Kaluza \cite{3}. He achieved his goal by 
adding an extra, fifth dimension to the visible four-dimenshional 
space-time and interpreted the $\mu 5$ $(\mu = 1,2,3,4)$ components as 
the electromagnetic vector potential. Afterwards  Oscar Klein \cite{4} 
suggested that this fifth dimension has a peculiar, periodic  topology, 
and, therefore, he "compactified" it to a  circles of small radii, 
attached to each point of our visible world.

    The fruitful Kaluza's idea  revivaled in 1980's \cite{5} and, more 
recently, with the advent of theories, known under the generic name 
"M theory" \cite{6}, which inevitabely lead to consideration of ten 
or eleven  (space)  dimensions. At last, at the end of preceeding 
century, interest to the extra dimentions greatly arose, because, on the 
one hand these theories may provide a natural solution of hierarchy 
problem and on the other hand give rich phenomenology for the 
collider physics \cite{2}. 
The idea is, that while all the usual particles live in our 
3-dimensional world (brane), gravitons are ubiquitous and live in 
additional dimensions (bulk) also. This explains why the Newtonian constant
is so small - gravitons waste their strength smearing it onto the extra
dimensions. Potentially, gravitational interaction can be 
enhanced by Kaluza-Klein excitations of gravitons up to order of 
electroweak one. How does this mechanism work see, for instance, in
the recent reviews \cite{7}.
According to concept outlined above the gravitational and electromagnetic 
interactions can be considered on the equal foot and influence one of 
them to another might be perceptible.
   Therefore, we can now speak about electrogravitational processes
hoping to observe them at the laboratory. Many such processes
are considered already \cite{8}, but in what follows the particular  
accent will be done on the mass dependence of the corresponding cross sections. 
Two types of particles (both are yet to be discovered) are related closely
to the masses: they are  Higgs bosons and gravitons. So, it is not unexpected, 
that some similarity might be loomed  between behaviours of these two particle 
types. Let us then first recall some features of the Higgs boson
production amplitudes.  
\section{Preliminaries}

Consider the so called Bj\"orken process,
 $\ell^+ + \ell^- \, \to \, Z + $ higgs $ \equiv H^0.$ 
\begin{figure}[htbp]
\centerline{\epsfig{file=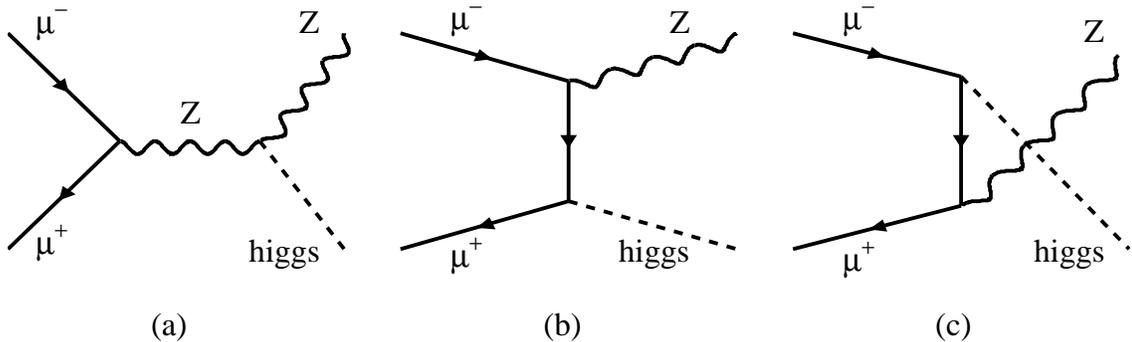,width=16cm,bbllx=0pt,bblly=0pt,bburx=397pt,bbury=142pt}}
\vspace{10pt}
\caption{Diagrams contributing to the process 
$ \ell^+ + \ell^-\rightarrow Z + H^0 $ are shown. Notations are
selfevident.}
\label{fig1}
\end{figure}
In calculation of the cross section for it one usually restricts 
to the consideration of the  graph (a) from  Fig.1 only. So do
we, but let us take into account masses of the initial fermions (muons, 
to be definite). 
Asymptotics of this process at $\sqrt s \rightarrow \infty$ is as follows 

\begin{eqnarray}
\sigma^{(c)}_{as}(\mu^+ \mu^- \, \to \, Z H^0)|_{m_{\mu}\neq0}
 ~=~ \frac{2\pi~\alpha^2 g^2_A}
{\sin^4(2\theta_{\cal E \cal W})}\frac{m^2_\mu}{m^4_Z},
\end{eqnarray}
where  $g^2_A$ is the constant of weak axial-vector coupling of 
the Z boson to muon, $\theta_{\cal E \cal W}$ is the angle of electroweak
mixing, and all other notations are selfevident.
It is easily to see that this result contradicts the unitarity
condition. Indeed, the cross-section obtained is azimuthal angle
independent. It means, that the scattering process occurs in the s-wave.
But the s-wave amplitudes  must satisfy the inequality
$$\sigma_{J=0} ~\leq~1/s,$$ while one sees from Eq.(1), that it is
constant (equal to $\cong 1.2 \cdot10^{-2}~fb$). In the expression above 
$J$ is the angular momentum of scattering,

In order to avoid the contradiction we include  two
more amplitudes in the calculation corresponding to the diagrams (b) and (c) 
of Fig.1. Cross section due to the sum of these two amplitudes is angle 
independent and is equal to $\cong 1.2 \cdot10^{-2}~fb$ again, but 
the interference between them and that of (a) results in 
$~~\cong-2.4 \cdot10^{-2}~fb$, completely eliminating confusion \cite{9}.
 This remarkable fine tuning reflects fundamental 
features of the underlying Standart Model \cite{10}. Namely, it is related 
to the renormalizibility of the latter, because the asymptotic behaviour 
of the tree amplitudes reflects the behaviour of the loop integrals with 
respect to their limits.

\section{Graviton-photon annihilation of $\ell^+\ell^-$ pair}

Now, let's turn to the "analogous" process 
$ \ell^+ + \ell^-\rightarrow \gamma + \cal G $, to which diagrams 
on the Fig. 2 correspond. 
\begin{figure}[htbp]
\centerline{\epsfig{file=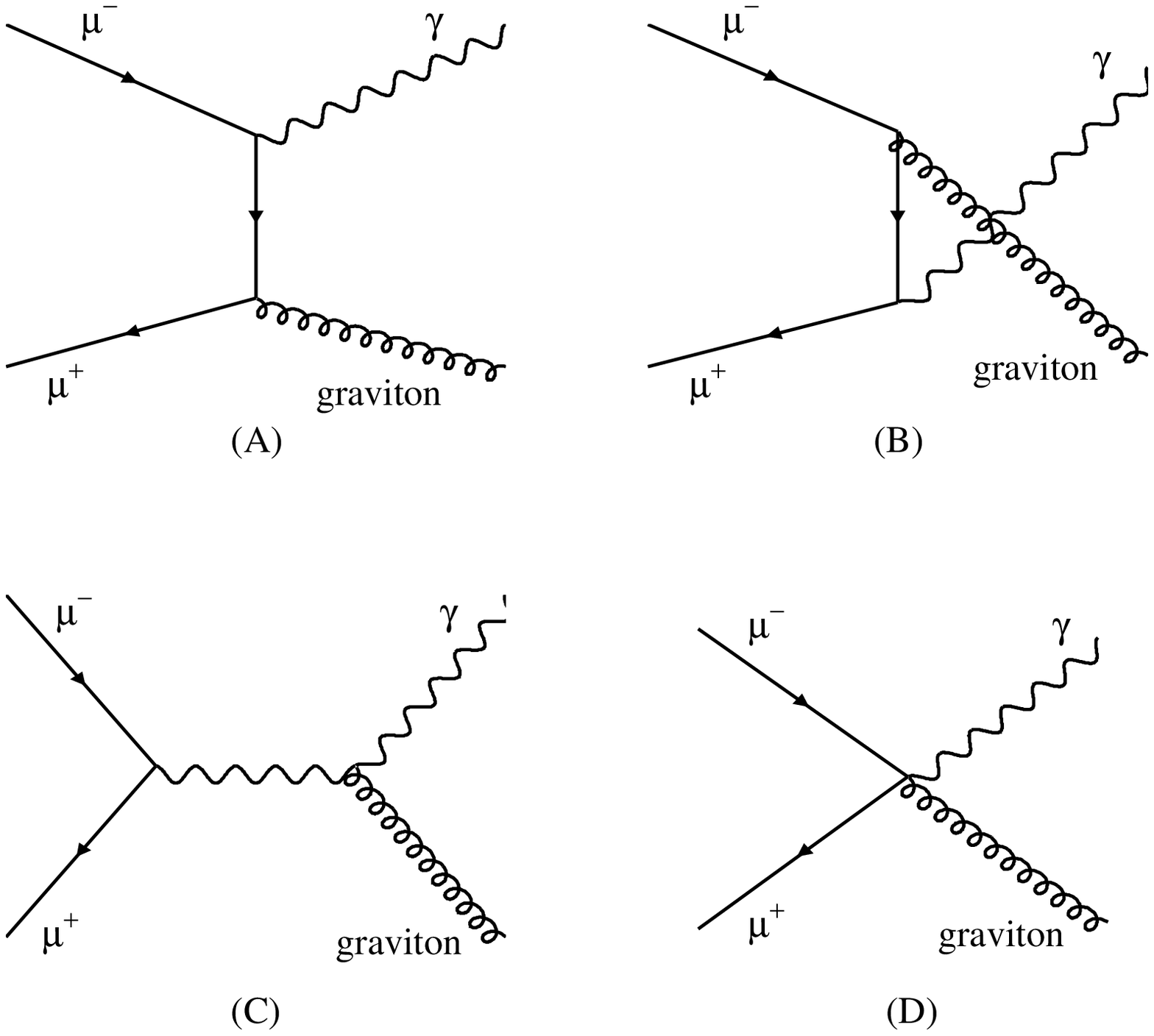,width=0.68\textwidth,bbllx=0pt,bblly=0pt,bburx=482pt,bbury=425pt}}
\vspace{10pt}
\caption{Diagrams contributing to the process
$ \ell^+ + \ell^-\rightarrow \gamma + \cal G $ are shown. Notations are
selfevident.}
\label{fig2}
\end{figure}
They are labeled by capital letters: 
(A) + (B) are {\bf Compton like}, (C) is the {\bf photon exchange} and
(D) is the {\bf contact} one, for future references.

In what follows we consider several cases with respect to the mass 
values of participating particles.

\subsection{Massless graviton, the lepton mass values are 
neglected}
The cross section is given in the differential form as

\begin{eqnarray}
\frac{d\sigma}{d\cos \vartheta_{\gamma}}
= \frac{\pi}{2}\alpha G^{enh}_{N}\biggl(1 + \cos^2\vartheta_{\gamma}\biggr),
\end{eqnarray}
where $\vartheta_{\gamma}$ is the emitting angle of 
the photon with respect to the negative lepton beam and $G^{enh}_{N}$ 
is now the constant of graviton coupling to the matter fields enhanced by the 
virtue of the Kaluza-Klein mechanism.

If, however, we calculate the cross section, corresponding to the
diagram (C)  of Fig.2 only, we obtain the same result as that of Eq.(2).
This is a consequence of the Ward identity.
For the illustration, let us write down only the contributions
of the Compton like diagrams (A) and (B).
Squaring the sum of (A) and (B) amplitudes, 
we obtain the following expression
\begin{eqnarray}
\frac{"d\sigma"^{Compton \; like}}{d\cos \vartheta_{\gamma}} 
= -\frac{5}{2}\pi\alpha G_N^{enh}  
\end{eqnarray}
Note, that this expression is constant and scattering angle
independent. 
Therefore, in the full expression for
the cross section the interference term appears:
\begin{eqnarray}
2\Re e \Bigl( {\cal M}_A + {\cal M}_B \Bigr)  \Bigl( {\cal M}_C + 
{\cal M}_D \Bigr) ^* + 2\Re e {\cal M}_C{\cal M}_D ^*.
\end{eqnarray}
This term completely compensates the contribution of the Compton like diagrams (A) and (B).
Note that after summing over the lepton polarizations
the squared (D) diagram vanishes:
\begin{eqnarray}
\sum_{spins}|{\cal M}_D|^2 = 0 \;\;\; .
\end{eqnarray}

It is expedient to mention the paper \cite{11}, where the fact of
mutual cancellation between contributions of diagrams,
describing the gravity in the strong electromagnetic field, was also noted.

\subsection{Massless graviton, the lepton mass values are not
neglected}

First, let us calculate that piece of cross section, which corresponds
to the {\bf photon exchange} diagram only in order to see if the case
of preceeding subsection is repeated. The result is as follows

\begin{eqnarray}
\frac{d\sigma^{(C)}}{d\cos \theta_{\gamma}}
= \frac{\pi}{2}\alpha G^{enh}_{N}\biggl(2-\beta^2 \sin^2\theta_{\gamma}\biggr),
\end{eqnarray}
while the cross section corresponding to the full set of diagrams is given 
in the differential form as
\begin{figure}[htbp]
\centerline{\epsfig{file=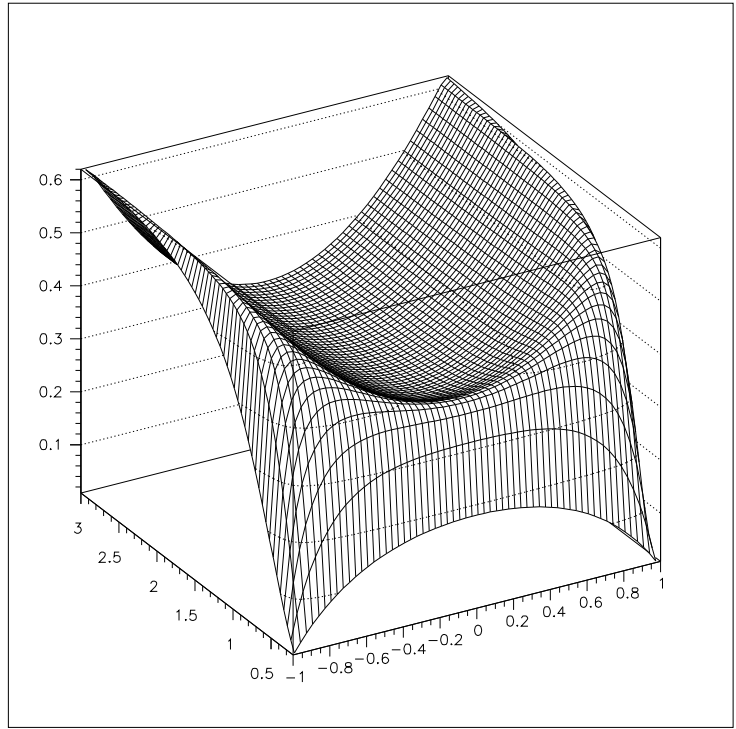,width=0.6\textwidth,bbllx=0pt,bblly=0pt,bburx=567pt,bbury=567pt}}
\vspace{10pt}
\caption{Angular distribution versus energy dependence of the 
process under study is shown.} 
\label{fig3}
\end{figure}
\begin{eqnarray}
\frac{d\sigma}{d\cos \vartheta_{\gamma}}
=\frac{1}{2}\pi \alpha G^{enh}_{N}\beta
\frac{1 + 2 \beta^2\sin^2\vartheta_{\gamma} - 
\beta^4 \bigl(1 + \sin^4\vartheta_{\gamma}\bigr)}
{(1 - \beta^2\cos^2\vartheta_{\gamma})^2}\sin^2\vartheta_{\gamma},
\end{eqnarray}
where\,\,$\beta=\sqrt{1-\frac{4m_{\ell}^2}{s}}$,  ~ $m_{\ell}$ ~
is the mass of initial lepton.

Evidently, the expression of Eq.(6) and that of Eq.(7) look utterly unalike -
the marvellous cancellation of the preceeding subsection disappeared
due to presence of lepton masses. 
Explicitly its behaviour with respect to the scattering angle and 
to the center of mass energy of collision is depicted on the Fig.3.
Being zero at the threshold it growths with energy and quickly
goes to plateau. On this picture both variable change in the following
limits: $ -1 \leq cos\vartheta_{\gamma} \leq +1,$ and $ 2m_{\mu}\simeq
0.212~ GeV \leq \sqrt{s} 
\leq 3~ GeV.$  Units for the cross section values are arbitrary.

Integrating the expression (7) over $cos \vartheta_{\gamma}$ in 
the limits above we obtain 

\begin{eqnarray}
\sigma_{tot} = \frac{4}{3}
\pi \alpha G^{enh}_{N}
\Biggl\{(1+8\frac{m_{\ell}^2}{s})(1-3 {\cal L}) + 12\frac{m_{\ell}^4}{s^2}
(1+2 {\cal L})\Biggr\}~~,
\end{eqnarray}
where $ {\cal L} = \frac{m_{\ell}^2}{s}\frac{1}{\beta}
\ln\frac{1+\beta}{1-\beta}$.

\subsection{Massive graviton, the lepton mass values are 
neglected} 
When the lepton mass is negligible and 
the graviton mass $m_{\cal G}$  is kept finite, the expression for 
the cross  section looks as 

\vspace{0.3cm}
\begin{eqnarray}
\frac{d\sigma}{d\cos\vartheta_{\gamma}}
= \frac{\pi}{2} \frac{\alpha G^{enh}_{N}}{(1-\frac{m_{\cal G}^2}{s})^2}
\Biggl\{1-(\frac{m_{\cal G}^2}{s})^5 
-(1-\frac{m_{\cal G}^2}{s})^3\cos^2 \vartheta_{\gamma}
\biggl[3\frac{m_{\cal G}^2}{s} + (1-\frac{m_{\cal G}^2}{s})^2\cos^2 
\vartheta_{\gamma} \biggr]\Biggr\}\frac{1}{\sin^2 \vartheta_{\gamma}}
\end{eqnarray}

   The remarkable feature of this expression is the fact, that it
has smooth limit at $m_{\cal G}\rightarrow  0.$ Namely,
in this limit we obtain the expression of the Eq.(2). Thus, the so called
van Dam - Veltman - Zakharov discontinuty \cite{12} is absent
in the physical quantity - cross section, although in deriving
expressions of Eq.(2) and that of Eq.(9) we have used projection
operators for the massless and massive gravitons, for which such 
a discontinuty presents explicitly. 

Now, the cross-section of the Eq.(9) is evidently singular
with respect to the scattering angle. This is consequence of neglecting
the initial state masses. The analytical expression for the 
cross section with the initial muon masses taken into account is
too cumbersome, so we will treat this case numerically in the next
subsection.

\subsection{Massive graviton, the lepton mass values are 
not neglected}
In this subsection we present the results of the numerical study
of the process under investigation with masses of participating
particles kept finite ( apart from the zero mass value for the photon,
obviously).

 Results of this analytical
computation of the matrix element squared with the aid of HECAS
program \cite{13} and Monte Carlo phase space integration are very 
different from that of the massless case and are  as  follows.

We consider only the one particular case, when 
$m^2_{\cal G}/s \rightarrow 1,$ i.e. $E_{\gamma}\rightarrow 0.$
It appears, that in this case the cross section behaviour is completely 
determined by the {\bf Compton like} diagrams only,
$$\sigma^{all\;diagrams} \equiv \sigma^{tot}\cong \sigma^{"Compton\;like"}$$
and there is no infrared cut-off by gravitation, as in the case
of massless graviton, i.e.
\begin{eqnarray}
\sigma^{tot}|_{m^2_{\cal G} \rightarrow s} \rightarrow \infty .
\end{eqnarray}
At the same time contributions from the {\bf photon exchange} diagram
and the {\bf contact} one tend to go to the zero in this limit, 
$\sigma^{(C\;+\;D)}|_{m^2_{\cal G} \rightarrow s} \rightarrow 0,$
while the interference between two subsets, namely, between
{\bf Compton like} and the sum of {\bf photon exchange} and {\bf contact} one 
remains constant and negative, $\sigma^{(A\;+\;B)(C\;+\;D)}|_{m^2_{\cal G} 
\rightarrow s} \simeq const,$ which , however, has no 
strong influence on the whole cross section behaviour in view of
Eq.(10). So, we were not able to draw any definite conclusion from the case
under consideration as opposite to the case of massless graviton.
 Nevertheless, questions raised
in this talk, deserve future study, because not all possibilities 
are yet considered. To this end we need the fully analytical expression 
for the cross section in the most general case, i.e. when all masses 
of the participating particles except photon,  are taken into account.
We hope to do this job in the nearest future.

\section{Conclusion}
   By a simple example of the process $ \ell^+ + \ell^-\rightarrow 
\gamma + \cal G$ the interplay between electromagnetic and gravitational 
interactions is considered. Some striking features of the amplitudes  
behaviour are revealed:
\begin{itemize} 
\item
in the case of massles graviton and in the limit of neglecting the
initial state mass values the cross section is completely determined
by the {\bf photon exchange} diagram only (diagram (C) of Fig.(2));
\item
the phenomenon disappears, when masses of initial states are taken
into account;
\item
in the case of massive graviton the behaviour of amplitudes is very 
different from that considered above.

\end{itemize}

\section{Acknoledgments}
I would like to thank Prof. Y.S.Kim for invitation me to participate in the 
work of Symposium and Mrs. Mary Ridgell for excellent job. Also, I am 
indebted to Drs. O.S.Pavlova and S.K.Stepanyan for useful information
they delivered to me and to Prof. D.V.Gal'tsov for making 
me aware of  Ref.\cite{11}.

\end{document}